\documentclass{PoS}
\usepackage{graphicx}
\usepackage{natbib}
\usepackage{journals}
\usepackage{deluxetable}

\title{Looking to the future: using IR interferometry to study microquasars}

\ShortTitle{Using IR interferometry to study microquasars}

\author{\speaker{Sera Markoff}\\
        Astronomical Institute ``Anton Pannekoek''\\
	University of Amsterdam\\
        E-mail: \email{s.b.markoff@uva.nl}}

\abstract{Infrared interferometry is currently in a rapid development
  phase, with new instrumentation soon achieving milliarcsecond
  spatial resolutions for faint sources and astrometry on the order of 10 microarcseconds.   For jet studies in particular, the next generation of instruments will bring us closer to the event horizon of supermassive black holes such as Sgr A*, and the region where jet launching must occur.  But a new possibility to study microquasars in general and jet physics in particular may also arise, using techniques similar to those employed for finding faint exoplanets around stars.  The compact, steady jets observed in the hard state of X-ray binaries display a flat/inverted spectrum from the lower radio wavelengths up through at least the far-IR band.  Somewhere above this band, a turnover is predicted where the jets become optically thin, revealing a power-law spectrum.  This break may have been observed directly in GX339-4, but in most sources such a feature is likely hidden under bright emission from the stellar companion or accretion disk components.  Detecting the exact location of this break would provide a new constraint on our models of jet physics, since the break frequency is dependent on the total power, as well as internal density and magnetic field.  Furthermore, knowing the break location combined with the spectral index of the power-law would help constrain the amount of synchrotron emission contributed by the jets to the X-ray bands.  Along with a summary of some potential observations requiring less optimal instrumental specifications, I will discuss a technique which may enable us to discern the jet break, and the chances of success based on theoretical models applied to some potential target sources.  }

\FullConference{Seventh Microquasar Workshop\\
		1-5 September 2008\\
		Fo\c ca, Izmir, Turkey}

\begin{document}

\section{Introduction}

Interferometry is not a new technique, having been developed many
years ago in the radio bands culminating in, e.g., Very Long Baseline
Interferometry (VLBI) astronomy and the Plateau de Bure Interferometry
(PdBI) at IRAM.  However as one goes to higher frequencies such as the
submillimeter, infrared (IR) and optical bands, correcting for
visibilities because of atmospheric effects becomes more difficult, and
to reduce these effects in general one cannot integrate as long on any
given source.  As a consequence, only very bright sources can
typically be observed at the moment (e.g., Altair's surface at V and
H-band photometric magnitudes of $\lesssim1$; \citealt{Monnieretal2007}).
However the gain in spatial precision at higher frequency is quite
significant, as the spatial resolution is defined by the fringe
spacing divided by the baseline between antennae, $\lambda/B$.
Typical resolutions in the optical are therefore more on the order of
milliarcseconds (mas) rather than arcseconds in the case of PdBI, for
instance.  Obviously different wavebands are important for different
types of emission, but for many astrophysical systems optical/infrared (OIR)
interferometry is the only method to resolve individual system
components.  Most often we hear about its potential as a technique for
detecting new exoplanets around distant stars, but here I will
consider how it may be useful also for studying microquasars.

One of the main differences between radio and OIR interferometry is
that one can perform phase referencing fairly easily in the radio
bands.  By pointing the telescope occasionally at a source with known
phase, the phase of the observed target can be calibrated.  However in
the IR/optical this is more challenging, because at the same time as
being more sensitive to atmospheric turbulence, the field of view is
generally much smaller and thus the probability of having a reference
star goes down.

There are several major facilities currently conducting OIR
interferometry, such as the Very Large Telescope Interferometer (VLTI)
in Paranal, Chile, the Keck Interferometer on Mauna Kea, and the Center
for High Angular Resolution Astrophysics (CHARA) on Mt. Wilson in
California.  In this talk I have focused on the VLTI, because of some new
instruments in development that I think will be potentially
interesting for microquasar studies.  The idea is to explore whether
these new instruments will enable us to find better constraints on jet
physics than currently possible with non-interferometric observations.

\section{Quick overview of the technique}

I am by no means an expert in IR interferometry (far from it
actually), but I became interested in its potential over the years
from attending meetings focusing on the Galactic Center. Adaptive
optics have pushed the field to the point where we can now observe the
orbits of individual stars around our Galaxy's central supermassive
black hole, Sgr A*.  These studies, conducted with both the VLT and
Keck \citep[e.g.][]{Ghezetal2000,Genzeletal2000} allowed the
determination of orbital parameters, as well as the detection of Sgr
A* itself for the first time in the IR, thus also enabling the discovery of IR flares in tandem with those found in the X-rays \citep{Baganoffetal2003,Genzeletal2003,Ghezetal2004}.  

Now that orbits have been charted for many stars in the inner parsec,
the envelope can be pushed even further by searching for general
relativistic effects occurring very close to Sgr A*.  There are
predictions that some stars may have periastrons even closer than
those already observed, but are currently too faint to detect with
current interferometers.  Similarly, the hint of substructure reported
in some of the Sgr A* flares \citep{Genzeletal2003} has peaked
interest in the possibility that the source of the flares, such as a
hotspot, could be followed while orbiting the black hole near its last
stable orbit.  In order to achieve the necessary sensitivity and
precision astrometry for these observations, an instrument for the
VLTI called GRAVITY is being developed, currently in preliminary
design phase \citep{Eisenhaueretal2008}.  By using both adaptive
optics (AO) via a guide star within 1', and a phase reference star
within the 2'' field of view, GRAVITY is designed to achieve 10
$\mu$as astrometry with the 8m VLT Unit Telescopes (UTs).  By tracking
the fringes (the interference pattern) on the bright reference object
one can actively stabilize the fringes of a second object. This allows
longer integration times of the fringes of the second object, which
can therefore be much fainter than the reference object. To achieve
the 10 $\mu$as astrometric accuracy, a laser must be used to measure
the path lengths in the system to nm accuracy and thus stabilize the
fringes. The VLT point-spread-function for imaging is $\sim$mas, thus the superb
stability of GRAVITY will allow one to determine motions of objects to
precisions $\sim100$ times better than the measurement of their
structure.

Although I think that GRAVITY will be very exciting, and will be the
only way to see very faint sources, it will likely not be operational
at the VLT until 2013.  In the meantime there is another VLT
instrument, the Phase-Referenced Imaging and Micro-arcsecond
Astrometry (PRIMA) facility \citep{Delplanckeetal2000}, that can serve
as the testbed for this idea, and in fact is probably already suitable
for the needs of a typical microquasar, which is not actually very
faint in the IR during outbursts.  PRIMA has already seen first
fringes in September 2008 and is currently being tested on the
Auxiliary Telescopes (ATs, 1.8 m) and eventually will be available for
use on the UTs.  The fringe tracking on the
brighter object (needs to be $K\lesssim 10$ on UTs, $K\lesssim 8$ on
ATs) allows stabilization of the fringes on a fainter object, as
described above, and for this instrument such a technique allows
synthesis mapping using, e.g., (M)IR instruments MIDI \citep{Leinertetal2003}
and AMBER (Astronomical Multi-BEam Recombiner;
\citealt{Petrovetal2007}) potentially down to $K\sim12-14$ on the ATs
and $K\sim 16$ for the UTs.  Like GRAVITY, 10 $\mu$as astrometry at
least on the ATs will be possible, but the phase reference star can be
as far as 20'' because of the larger instrumental field of
view.

\section{Microarcsecond astrometry of microquasars: the idea}

How can we use this for microquasar studies?  Most microquasars have a binary separation falling in the range of 10--100 $\mu$as, clearly not resolvable with the VLT.  I will focus instead on the potential for studies using astrometry.  

In the most ``classical'' astrometric observation, we can search for
the wobble of the companion star (as measured with respect to the
reference star) due to the compact object primary, and use this to
constrain the binary parameters of a system.  This is exactly the
technique used to find exoplanets, which are typically orders of
magnitude dimmer than a jet would be during outburst, and much lighter
than, e.g., a stellar mass black hole.  Such an application alone
would be very valuable, given how few sources have well-determined
mass functions and orbits.  For both high mass X-ray binaries (HMXBs)
and low-mass X-ray binaries (LMXBs), this should be achievable if
another $K\sim 8-10$ magnitude source is within the 20'' field of
view.

A similarly ``classical'' PRIMA + AMBER observation (that has nothing
to do with astrometry, however) would be to search for a
drop in visibility as the baseline is varied, which would suggest that
the source has been resolved out.  This technique provides a way to
estimate the binary separation without actually imaging the two
components.  AMBER is sensitive to changes in visibility of
$\sim0.2$\%, and this could be a way to obtain system physical
parameters below the nominal mas resolution of the VLT.

I consider these as the baseline applications of IR interferometry for
microquasars/XRBs, and something we should consider proposing for routinely
once the facilities are available.

There are, however, a few additional potential applications that will
be very interesting if they actually turn out to be feasible.  One
possibility is to measure the change in closure phase (see, e.g.,
\citealt{PearsonReadhead1984}) as a probe of asymmetries that could
tell us something more about the presence or lack of jets during
outbursts.  I have not had time to explore the feasibility for this
case study, though, and will not discuss it further.  I will focus
instead on two applications that both rely on the detection of a shift
in phase due to changing intrinsic components rather than a physical
wobble due to the orbital motion.  Understanding the orbital solution
via the ``wobble'' technique or otherwise will be necessary,
especially in cases where we may want to co-add several orbits.  

The light centroid is basically the equivalent of a center of mass, with flux substituted in for the mass terms.  Imagine if a system were viewed perfectly edge on, when it was magically not orbiting.  The primary and secondary stars will not be resolved, and their combined emission will basically look like a single point source.  If the ratio of the primary and secondary fluxes changes either as a function of time or frequency enough, and the binary separation is large enough (basically 10 $\mu$as or more), PRIMA, (in one case necessarily combined with MIDI and/or AMBER) will be able to detect this shift as a change in relative measured phase.  For this to work, the dynamic range in the system cannot be too large, because the jet flux must be a significant fraction of the companion star's in order to have an observable affect on the centroid.

\begin{figure*}
\centerline{\includegraphics[width=0.5\textwidth]{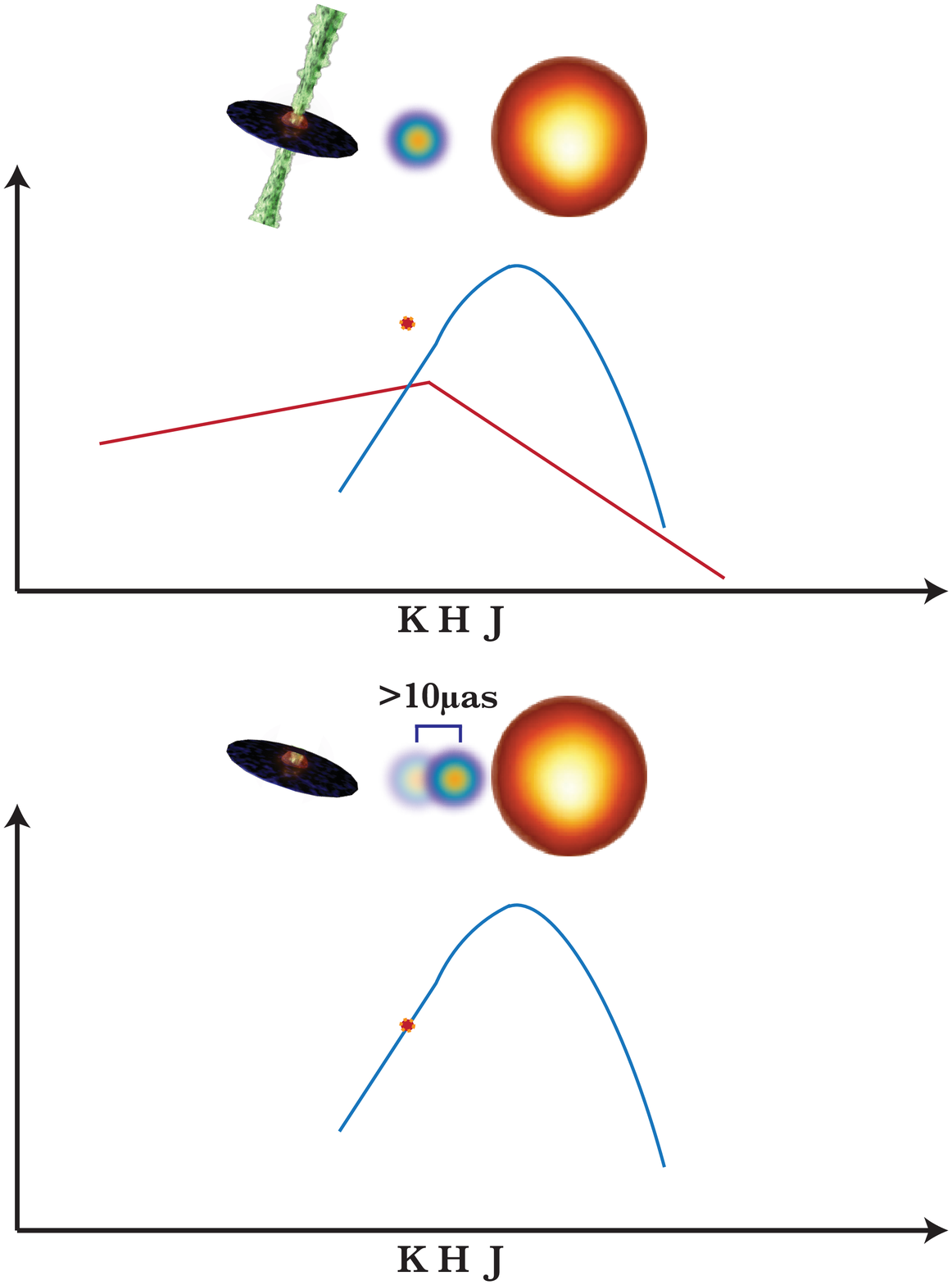}\hfill\includegraphics[width=0.5\textwidth]{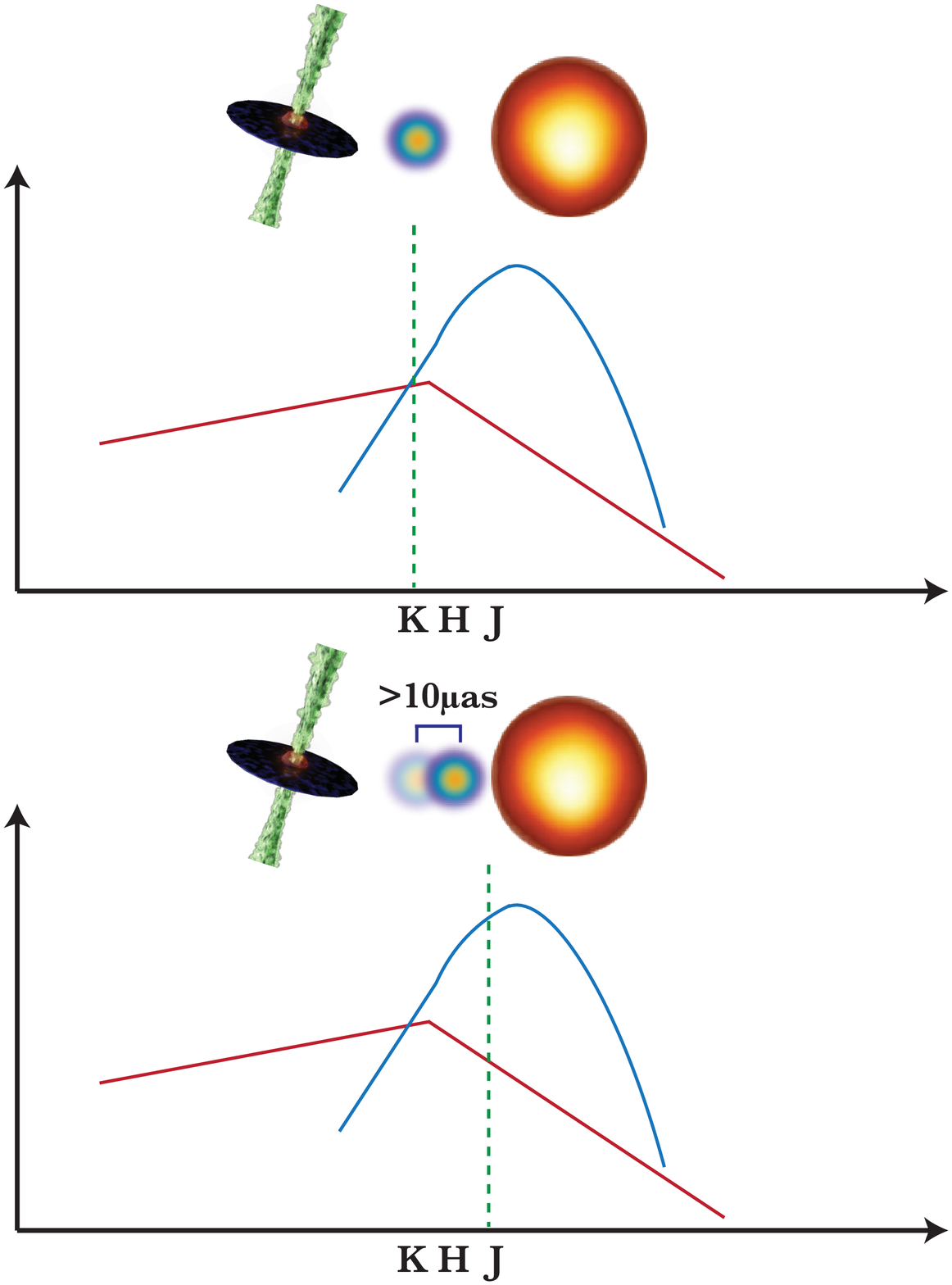}}
\caption{\textsl{Left panels:} Schematic of the centroid shift expected during state changes, due to the disappearance of the jet infrared synchrotron emission.  This type of measurement could help constrain the binary separation. \textsl{Right panels:} Schematic of the centroid shift expected from the relative change in jet and stellar fluxes between K and J bands.  This type of measurement could also constrain the binary separation as well as the location of the jet synchrotron break frequency, and potentially the spectral index of the jet emission as well. }
\label{centroid}
\end{figure*}

The first idea is to look for such a change at a \textit{fixed} frequency by comparing the centroid location between the hard (jets on) and soft (jets off) states, as illustrated in the left top and bottom panels of Fig.~\ref{centroid}.  A detection of a shift in the centroid before/after the hard/soft transition would first of all allow us to estimate the binary separation if the jet flux is known (or can be extrapolated) from non-interferometric observations.  At the very least, better limits on the jet flux and binary separation could be obtained. Note that this detection requires the presence of a bright enough reference star within 20'', since the shift is measured with respect to the reference star fringes.

I find the second scenario even more interesting, because it does not
require the presence of a reference star and so potentially could be
conducted for sources with $K>9$ using AMBER alone without the need for PRIMA, however using PRIMA would lower the detection threshold significantly.  In all cases, it is only feasible for sources with a large angular separation, thus likely high mass and nearby systems.  As illustrated in the right top and bottom panels of Fig.~\ref{centroid}, the shift in the centroid can also provide a measure of the spectral energy distribution (SED) of the jet in a region normally blocked by the bright companion emission.  Assuming that the stellar flux is well characterized in this regime by the Rayleigh-Jeans tail of a black body, the presence of a break in the jet spectrum could be constrained.  If this technique is determined to be robust, eventually the location of the break as a function of other variables such as accretion power could be monitored as a part of multiwavelength campaigns.  

Why would this be interesting?  First of all, because the optically
thick synchrotron emission from the compact jets is flat or inverted,
the break frequency determines the total radiative power, placing a
lower limit on the total power of the jets.  Secondly, it could help
establish the extent to which synchrotron from the jets is
contributing to the X-ray emission during the hard state.  Particles
are clearly being accelerated in microquasar jets, as evidenced by the
power-law synchrotron emission detected during optically thin
outbursts.  If this acceleration begins at a defined location in the
jets, such as a collimation shock, or where the flow surpasses the
slow magnetosonic point and can develop such disturbances, the typical
values of the local particle density and magnetic field make it likely
that synchrotron emission extends into the X-ray band.  My colleagues
and I originally explored the most extreme scenario for XTE~J1118+480
in \cite{MarkoffFalckeFender2001}, where the X-rays were entirely
dominated by synchrotron emission.  More likely for most sources,
synchrotron contributes at a roughly $\sim10$\% level, with maximum
dominance in the soft band \citep{MarkoffNowakWilms2005}.  Such a scenario is shown in the righthand
panel of Fig.~\ref{sources}.  The extent of a synchrotron contribution
can be gauged by a stringent test of the ``IR coincidence'', as coined by
\cite{Nowaketal2005}.  The location of the break should be
strongly correlated with the soft X-ray flux if synchrotron is
significant, otherwise the line-up is just coincidence.  By combining IR
interferometry with X-ray monitoring, this issue could be quickly
resolved.

Similarly, simple jet scaling physics predicts a direct relation of
the synchrotron self-absorption break frequency on the mass and power
in the jets.  For the case of the jet power linearly depending on
$\dot{m}$, with a fixed partition of conserved particle energy density
and magnetic energy density, and a power-law of emitting particles
$p=2$, this scaling goes as
\begin{equation}
\nu_{jb}\propto M^{-1/3}\dot{m}^{2/3},
\end{equation}
where $\nu_{jb}$ is the jet break frequency and $M$ is the black hole mass \citep{FalckeBiermann1995,Markoffetal2003,HeinzSunyaev2003}.  Deviations from this scaling can help cast light on internal jet physics, especially if the slope after the break $\alpha=(p-1)/2$ can be measured.

\section{Microarcsecond astrometry of microquasars: some examples}

How feasible will this kind of measurement be?  This is difficult to estimate since at the moment we do not have many sources where the binary separation and SED are both well constrained.  I have taken two examples just to run some sample numbers and see how they come out.  The spacing of an interferometer's fringes on the sky is $\lambda/B$, analogous to the $\lambda/D$ resolution of a single optical dish.  The phase of the centroid of the jet and companion star can be calculated from:
\begin{equation}
\phi = 360^\circ \left(\frac{a}{\lambda/B}\right)\frac{I_j}{I_j + I_s},
\end{equation}
where $a$ is the binary separation and $I_j$ and $I_s$ are the
measured jet and star flux at a given frequency, respectively.  By
calculating the resulting shift in phase $\delta\phi$ corresponding to
the centroid shift either between the jet on and off states, or
between K relative to J band as can be measured with AMBER, we can
test whether such a shift would be detectable.  A phase shift on the
order of 1$^\circ$ corresponds to $\sim 10\mu$as on the sky, and thus
should be visible with PRIMA.  

\begin{figure*}
\centerline{\includegraphics[width=0.57\textwidth]{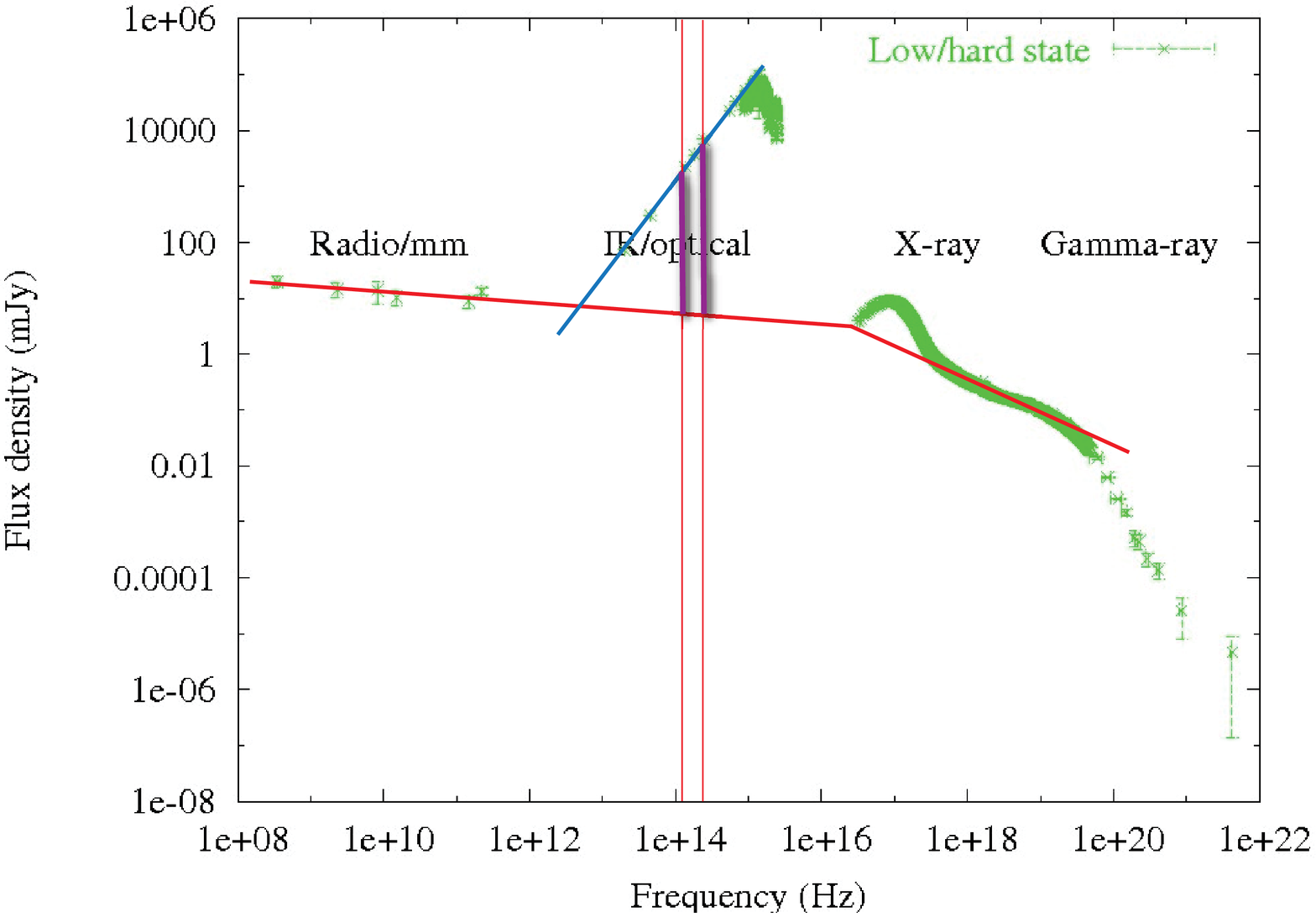}\hfill\includegraphics[width=0.43\textwidth]{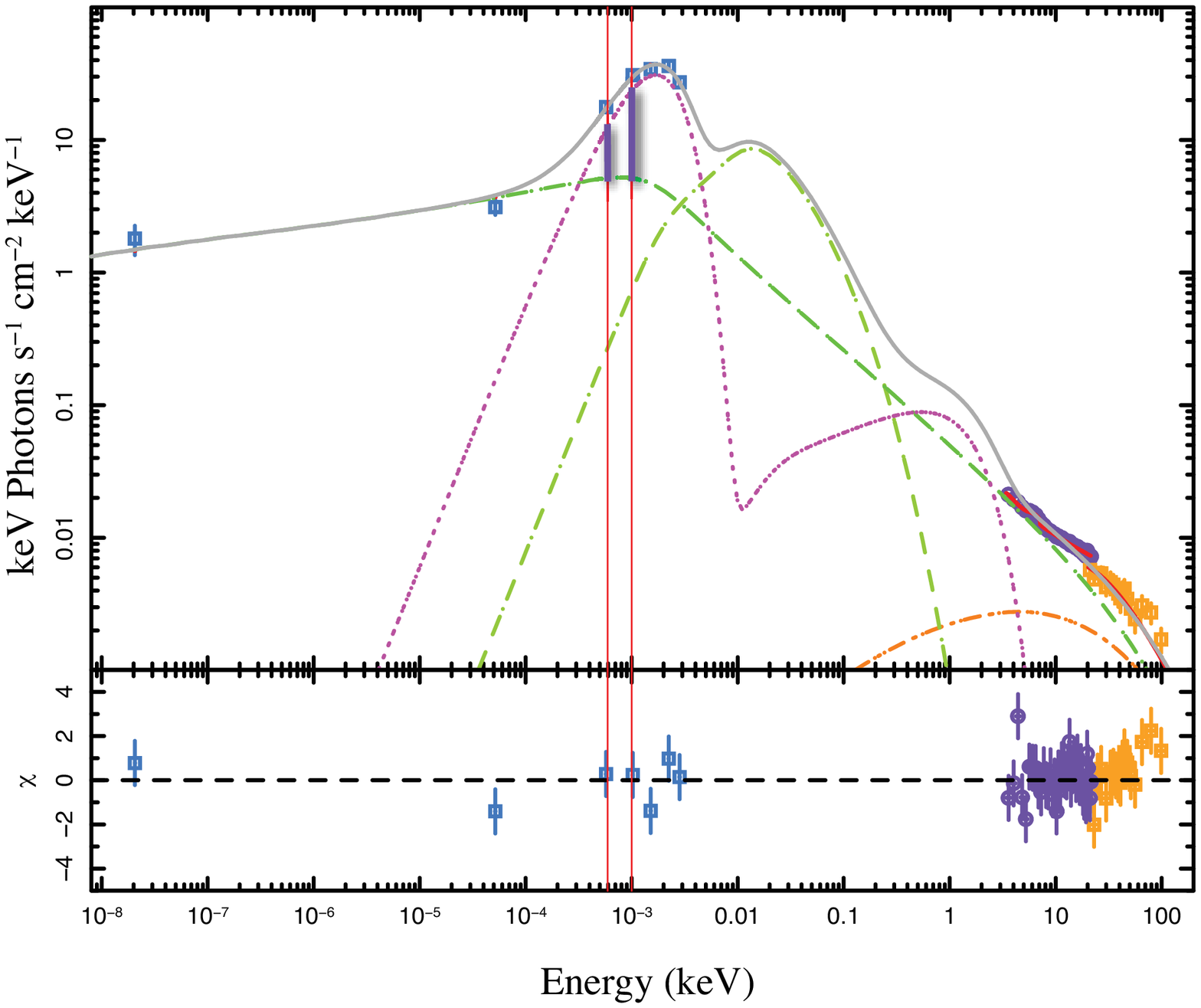}}
\caption{\textsl{Left:} Compiled average spectrum of Cyg X-1 in the hard state (Fender, Tigelaar, priv. comm.) illustrating the K and J bands (thin red lines) and the change in flux between the jet and disk contributions (thick purple lines) over this range. \textsl{Right:} Same labeling for a simultaneous hard state observation of GRO~J1655-40 from \cite{Migliarietal2007}.}
\label{sources}
\end{figure*}

In Figure~\ref{sources}, I show two representative spectra with lines
overdrawn to emphasize how I obtained the numbers in my estimates.
For each source I calculated the phase shift both for the ``jets
switching on/off'' scenario, and the steady state comparison between
J/K bands scenario.  The predicted shift is shown in Table 1. I used a
baseline of 100m to make a conservative estimate (200m is the maximum
VLT baseline for the ATs, lower for the UTs). 

\begin{deluxetable}{lcccc}
\tablewidth{0pt}
\tablecaption{Predicted phase shift for two example microquasars.  These numbers are model dependent but within an order of magnitude of accuracy.  \label{table}}
\tablehead{\colhead{Source} & \textbf{a ($\mathbf{\mu}$as)} &
  \colhead{J vs. K band} & \colhead{``Jets on/off, K''} & \colhead{``Jets on/off, J''}}
\startdata
Cyg X-1 & 80 & 0.02$^\circ$ & 0.04$^\circ$ & 0.07$^\circ$\\
GRO~J1655-40 & 40? & 0.005$^\circ$ & \textbf{0.99$^\mathbf{\circ}$} & \textbf{1.7$^\mathbf{\circ}$}\\
\enddata
\tablecomments{The binary separation for GRO~J1655-40 is not known, thus I have taken a value typical for LMXBs.  The assumed baseline was 100m.}
\end{deluxetable}

The first thing to note is that Cyg X-1 is not a viable source for
PRIMA + AMBER to detect the differential phase for either of the two
scenarios.  In reality it is not feasible for the VLT under any
condition because of its location in the northern sky, but I felt it
was a good ``canonical'' test source whose spectrum is relatively
well-understood.  The point is that this HMXB has such a bright
companion star that the contribution of the jet is too small to shift
the phase regardless of whether the jets are on or off.  

On the other hand, a luminous, nearby LMXB with a bright companion
like GRO~J1655-40 may be a good target.  First of all it is in the
correct hemisphere, but also the ``dynamic range'' is less; the jet
contributes a significant amount of the IR flux and thus its presence
effects the differential phase.  The values for both the stellar and
jet fluxes are dependent on the model, however the stellar
component is fairly well constrained by the IR/optical data so if a
single blackbody is a reasonable assumption, our estimation of the jet
contribution is relatively robust.  Because for this particular model
shown here $\nu_{jb}$ is above the observed frequency range,
the flux difference between J and K bands is not enough to cause a
detectable change in centroid position.  However, the jet contributes
enough flux that its absence would cause an observable shift in the
phase, for PRIMA (when up to full capability).

Considering the case of PRIMA+AMBER, if $\nu_{jb}$ occurs just before
the K band, the jet flux in the J band would decrease by about 20-25\%
and the resulting phase shift would be $0.02^\circ$.  This is still
too small to observe by quite a margin, but it is not impossible to
imagine a closer source with a steeper jet break where this could be
detectable.  It is also important to note that I used a conservative
100m in all estimates so far.  If the largest possible VLT baseline of
200m were used, all of the above numbers would be doubled, but the
source would need to be bright enough to use the ATs rather than the UTs.

\section{Discussion}

The idea of using IR interferometry to make astrometrical measurements of microquasars is new, and likely there are technical issues that I am underestimating.  But from the sample numbers I have run so far, it seems that there are several potential avenues to explore to help us better characterize, at the very least, the orbital parameters of these systems, and at best, obtain new constraints on the jet physics.  For systems where nothing is known about the mass function, we can look for a wobble in HMXBs where the IR spectrum is dominated entirely by the companion star in order to constrain the orbit and/or mass of the primary.  But for bright, nearby LMXBs, there seems to be a real possibility for new ways of detecting microquasar jets.  Desirable targets will be in the Galactic Plane or southern sky, nearby, and have large orbital separations.  If the orbit can be constrained and any resultant wobble factored out, we should be able to use astrometry to place limits on the jet contribution and possibly even constrain the location of the break frequency and post-break spectrum.  Not only will this allow us to place a hard lower limit the radiative power of the jets, it will also enable us to place strong upper limits on the jet synchrotron contribution to the X-ray bands, a point of considerable debate.  

The issue of accounting for the orbit may be a huge caveat.  Also it
is not clear to me how many of these sources will have a good
reference star within 20'', but I think that in many fields we are
interested in, such as the Galactic Plane, there should be enough
candidates.  Given that most systems of interest have K-band
magnitudes greater than 7, for the differential interferometric
measurements we will need to wait until PRIMA is up to full operation
on the VLT UTs.  This delay may work to our advantage, however, since
this year the first radio all-sky monitor is starting up with the
commissioning of the Low-Frequency Radio Array (LOFAR;
\citealt{Fenderetal2008}), anticipated for full operation by the end
of 2010.  We expect to detect hundreds of new transients, many of
which will fall into the swath of sky visible also with the VLT near
the Galactic Plane.  With multiwavelength followup, we will likely be
able to identify several new candidates that match our wishlist for
testing the OIR interferometric techniques described here.

\subsection*{Acknowledgments}

This talk and subsequent article would not have been possible without significant guidance from, and
discussions with, Frank Eisenhauer (MPE), Roberto Abuter (ESO), Walter Jaffe (U Leiden),
Thibaut Paumard (Paris Observatory, Meudon) and Rainer Sch\"odel (IAA
Granada). However, any mistakes are solely my own.

\end{document}